\documentclass[aps,prl,showpacs,twocolumn,amsmath,amssymb,superscriptaddress]{revtex4}

\usepackage{graphicx}
\usepackage{dcolumn}
\usepackage{bm}
\usepackage{times}

\begin{document}

\title{Contrasting Behavior of Carbon Nucleation in the Initial Stages of Graphene Epitaxial Growth on Stepped Metal Surfaces}

\author{Hua Chen}
    \affiliation{Department of Physics and Astronomy, University of Tennessee, Knoxville, TN 37996}
    \affiliation{Materials Science and Technology Division, Oak Ridge National Laboratory, Oak Ridge, TN 37831}

\author{Wenguang Zhu}
    \affiliation{Department of Physics and Astronomy, University of Tennessee, Knoxville, TN 37996}
    \affiliation{Materials Science and Technology Division, Oak Ridge National Laboratory, Oak Ridge, TN 37831}

\author{Zhenyu Zhang}
   \affiliation{Materials Science and Technology Division, Oak Ridge National Laboratory, Oak Ridge, TN 37831}
   \affiliation{Department of Physics and Astronomy, University of Tennessee, Knoxville, TN 37996}
\date{\today}

\begin{abstract}
Using first-principles calculations within density functional theory, we study the energetics and kinetics of carbon nucleation in the early stages of epitaxial graphene growth on three representative stepped metal surfaces: Ir(111), Ru(0001), and Cu(111). We find that on the flat surfaces of Ir(111) and Ru(0001), two carbon atoms repel each other, while they prefer to form a dimer on Cu(111). Moreover, the step edges on Ir and Ru surfaces cannot serve as effective trapping centers for single carbon adatoms, but can readily facilitate the formation of carbon dimers. These contrasting behaviors are attributed to the delicate competition between C-C bonding and C-metal bonding, and a simple generic principle is proposed to predict the nucleation sites of C adatoms on many other metal substrates with the C-metal bond strengths as the minimal inputs.
\end{abstract}

\pacs{68.35.Fx, 68.43.Bc, 68.43.Hn}

\maketitle

Since its first isolation, graphene has attracted rapidly growing research interest because of its various intriguing properties and potential applications in future electronics \cite{geimnm07,castrormp09}. However, a route towards scalable mass production of quality graphene for industrial use is still lacking. Among many newly developed techniques, epitaxial growth of graphene on metal surfaces offers a promising avenue \cite{wintterlinss09,marchiniprb07,sutternm08,corauxnl08,yuapl08,liscience09,kimnature09,loginovanjp08,loginovanjp09,mccartycarbon09,lacovigprl09,linl09,corauxnjp09,loginovaprb09,vangastelapl09}. Large size and good quality graphene samples have been prepared on various metal surfaces \cite{wintterlinss09,marchiniprb07,sutternm08,corauxnl08,yuapl08,liscience09,kimnature09}. The success in transferring the epitaxial graphene grown on Ni and Cu surfaces to insulating substrates makes this method even more attractive \cite{yuapl08,liscience09}. Additionally, various aspects about the growth mechanisms of graphene have been revealed in recent studies of representative carbon/metal systems. For example, the growth of graphene on Ir(111) and Ru(0001) substrates is fed by the supersaturated two-dimensional (2D) gas of carbon adatoms, and a multi-carbon cluster attachment mechanism has been proposed  \cite{loginovanjp08,loginovanjp09,mccartycarbon09}, with minimal effect of hydrogen \cite{loginovanjp09,lacovigprl09}. On a Cu substrate, graphene is found to grow through a surface adsorption process, while on Ni it is by carbon segregation or precipitation \cite{linl09}.

Despite these preliminary achievements, very little has been revealed about the growth kinetics, especially in the initial nucleation stages of carbon adatoms. Experimentally it has been found that carbon nucleation starts from the lower edges of steps on Ir(111) \cite{corauxnjp09} and Ru(0001) \cite{loginovanjp08} surfaces, but it is unclear why and to what extent nucleation at the step edges is preferred over terraces. Determination of nucleation sites is crucial in improving both the quality and quantity of epitaxial graphene. In the growth of graphene on Ru(0001), multiple nucleation on terraces can easily degrade the quality of graphene because defects will form at the interfaces of separately nucleated graphene islands \cite{loginovanjp09}. In graphene growth on Ir(111), the nucleation sites must not be too sparse, because otherwise rotated graphene domains are more likely to grow at the boundaries of the major phase of the islands that are aligned with the substrate \cite{loginovaprb09,vangastelapl09}. Quantity-wise, in order to eventually achieve mass production for industrial applications, it is more desirable for nucleation of graphene islands to take place over the entire substrate rather than only at the edges of preexisting steps. In light of these aspects, a general guiding principle of determining the nucleation sites on different substrates will be highly beneficial.

	In this Letter, we present a comparative study of the energetics and kinetics in the initial stages of epitaxial graphene growth on three representative stepped metal surfaces, using first-principles calculations within density functional theory (DFT). We find that, whereas the interaction between two adatoms is attractive on flat Cu(111), leading to easy ad-dimer formation, it becomes repulsive on flat Ir(111) and Ru(0001), making ad-dimer formation improbable. On the other hand, even though the steps on Ir(111) and Ru(0001) cannot serve as effective trapping centers for single carbon adatoms, such steps can readily facilitate the formation of carbon dimers at their lower edges. We rationalize these contrasting kinetic behaviors of carbon adatom diffusion and nucleation based on the delicate competition between the C-C bonding and C-metal bonding, and generalize this picture to predict the initial growth stages of graphene on different metal substrates.

	In our studies, we use the Vienna ab initio simulation package (VASP) \cite{kresseprb96} with PAW potentials \cite{blochlprb94} and the generalized gradient approximation (PBE-GGA) \cite{perdewprl96} for exchange-correlation potential. All the metal surfaces are modeled by a 6-layer slab, with atoms in the lower 3 layers fixed in their respective bulk positions. A $(2 \times 4)$ squared surface unit cell is used to describe the Ir(111) , Ru(0001) and Cu(111) surfaces. We use (322) and (332) surfaces to model the stepped Ir(111) and Cu(111) surfaces, which contains \{100\} (A-type) and \{111\} (B-type) microfacets, respectively. The stepped Ru(0001) surface is modeled by a vicinal surface with its normal along the $ \langle 0 \, \overline{1} \, 1 \, 10 \rangle$ direction, which contains alternating A- and B-type steps. All the terrace widths are $\sim$11-12 {\AA}. The k-point mesh used in the calculations is $1\times 3\times 1$ for stepped Ru(0001), and $3\times 3\times 1$ for all the other cases \cite{methfesselprb89}. We use the climbing image nudged elastic band (CINEB) method \cite{henkelmanjcp00} to determine the energy barriers of the various kinetic processes.

\begin{figure}
 \begin{center}
\includegraphics[width=3.4in]{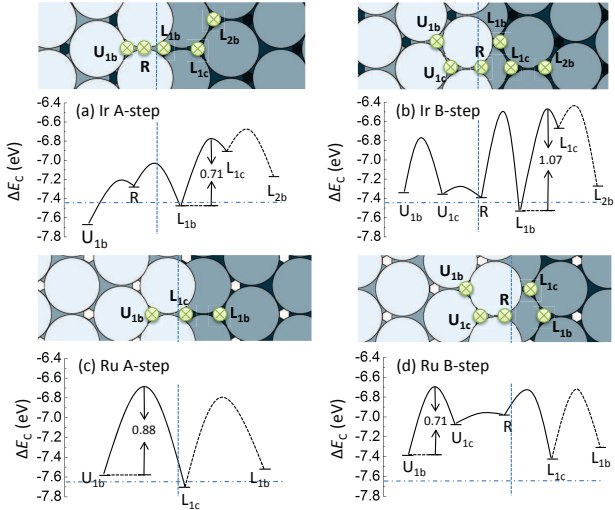}
 \end{center}
 \caption{\textbf{(Color online)} Top views of adsorption sites and binding energies of a C adatom around (a) Ir A-step, (b) Ir B-step , (c) Ru A-step, and (d) Ru B-step. The solid curves represent C diffusion profiles. The vertical dashed line represents the position of step edge, and the horizontal dot-dashed line indicates the C binding energy on flat surfaces. Definition of labels: U - upper terrace, L - lower terrace, R - ridge site where C only binds to atoms in the ridge of a step edge; b - hcp site, c - fcc site.}
 \label{figure1}
\end{figure}

	We first consider the adsorption and diffusion of isolated carbon atoms on flat metal surfaces. The most stable adsorption sites and the corresponding binding energies, defined by $\triangle E_{\mathrm{C}}=E_{\mathrm{C/subst}} -E_{\mathrm{C}}-E_{\mathrm{subst}}$, on Ir(111), Ru(0001) and Cu(111) are hcp (-7.44 eV), hcp (-7.66 eV) (in agreement with previous calculations \cite{loginovanjp08,loginovanjp09}), and subsurface interstitial (-5.66 eV), respectively. The stronger binding on the other two substrates and the weaker binding on Cu(111) are consistent with the \emph{d}-band model \cite{nilsson07}: on Cu(111), the C adatom mainly interacts with the free-electron like surface states of Cu, whose \emph{d}-shell is completely filled; whereas on Ir(111) or Ru(0001), the stronger binding originates from the hybridization between the \emph{sp} orbitals of carbon and the half-filled \emph{d}-band of the substrate. On Ir(111) and Ru(0001), the energy in the metastable fcc sites are 0.25 eV and 0.74 eV higher, respectively. On Cu(111), the metastable sites on the surface (fcc, hcp, bridge) are less stable than the subsurface interstitial sites by $\sim$0.6 eV. The surface diffusion barriers ($\varepsilon_a$) between a stable and the nearest metastable states are 0.75 eV, 0.87 eV and 0.66 eV on Ir(111), Ru(0001), and Cu(111), respectively.

	We next investigate the adsorption and diffusion of single C adatoms at step edges of Ir(111) and Ru(0001). The results for the case of Cu(111) will be reported in a separate work \cite{chen}, as will be explained later. As shown in Fig.~\ref{figure1}, the calculated binding energies at step edges are not much larger than those on flat surfaces. The same is true for the kinetic barriers. Considering the high growth temperatures in experiments ($\sim$1000 K), we arrive at the conclusion that the substrate steps do not serve as effective traps for single C adatoms. The absence of large step-crossing barriers and deep wells at step edges, in contrast to the traditional Ehrlich-Schwoebel (ES) picture \cite{ehrlichjcp66,schwoebeljap66,moprl08}, is attributed to the passivation of the least coordinated step edge atoms by the non-metal adatoms, leading to a large amount of energy gain \cite{feibelmanprl96}. Therefore, in the transition state where the adatom bonds to step edge atoms only, it more effectively passivates the edge atoms, resulting in a lower barrier. Similarly, because at lower step edges the substrate atoms are relatively over-bonded, C adatoms cannot gain much energy by having more neighbors \cite{feibelmanprl96}. Thus no extra deep potential well is present either.

\begin{figure}
 \begin{center}
\includegraphics[width=2.6 in]{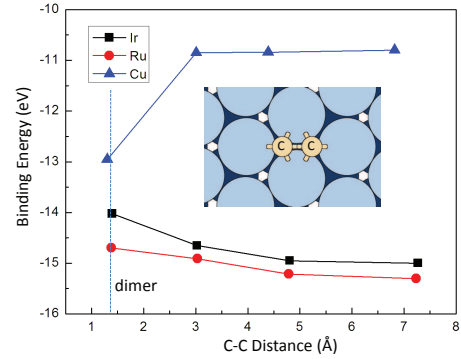}
 \end{center}
 \caption{\textbf{(Color online)} Binding energies of two C adatoms on flat metal surfaces as a function of their separation distance. Data points around the vertical dashed line correspond to the formation of C dimers. Inset shows the top view of a C dimer on a close-packed metal surface. Kinetic barriers are not shown.}
 \label{figure2}
\end{figure}

	Where should nucleation of carbon adatoms occur if they are not effectively trapped anywhere on the substrates of Ir and Ru? The above results indicate that knowing the behavior of non-interacting single carbon adatoms is insufficient to answer this question. Specifically, since carbon adatoms are known to form strong covalent bonds with one another when they nucleate to form graphene, it is necessary to take the carbon-carbon interaction into account. We therefore next study the formation of C dimers as the first step of nucleation on the metal substrates. Fig.~\ref{figure2} shows the trend of binding energies of two carbon adatoms on the flat metal surfaces, defined by $\triangle E_{\mathrm{2C}}=E_{\mathrm{2C/subst}} -2E_{\mathrm{C}}-E_{\mathrm{subst}}$, as a function of the separation distance. One can immediately notice that on Ir(111) and Ru(0001) the formation of C dimers is energetically unfavorable, but on Cu(111), dimers are much more stable than separate C adatoms by over 2 eV. Moreover, the energy barrier of forming a dimer for two neighboring C adatoms is only 0.32 eV on Cu(111), which is much smaller than those on Ir(111) (1.37 eV) and Ru(0001) (1.49 eV). These findings suggest that on Ir(111) and Ru(0001), C adatoms are mutually repulsive and cannot form dimers, whereas on Cu(111) they strongly attract each other, leading to the formation of dimers and larger islands.

\begin{figure}
 \begin{center}
\includegraphics[width=3.4in]{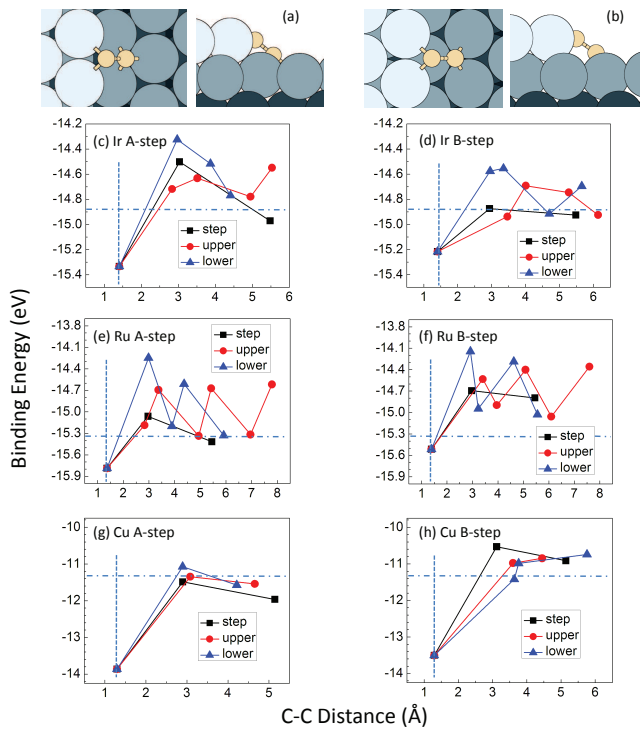}
 \end{center}
 \caption{\textbf{(Color online)} (a) and (b) Top- and side-view of the most stable configuration of a C dimer at the lower edge of a (a) A-step and (b) B-step. (c-h) Binding energies of two carbon adatoms with one C atom fixed at the lower step edge and another moving on the upper terrace, the lower terrace and along the lower step edge, respectively. Horizontal axis is their separation distance. Vertical dashed line in each panel shows where a C dimer is formed, and horizontal dot-dashed line shows the binding energy of two separate C adatoms on flat surfaces. }
 \label{figure3}
\end{figure}

\begin{table}
\caption{Binding energy difference between a C dimer at lower step edges ($E_{\mathrm{dimer/step}}$) and (1) two separate C adatoms on flat surfaces ($E_{\mathrm{2C/flat}}$) and (2) a C dimer on flat surfaces ($E_{\mathrm{dimer/flat}}$). (The unit is in eV).}
\label{table1}
\begin{ruledtabular}
\begin{tabular}{ccccccc}
Energy difference&\multicolumn{2}{c}{Ir}&\multicolumn{2}{c}{Ru}&\multicolumn{2}{c}{Cu}\\
Step&A&B&A&B&A&B\\
\hline
$E_{\mathrm{dimer/step}}-E_{\mathrm{dimer/flat}}$&-1.31&-1.19&-1.09&-0.68&-0.64&-0.28\\
$E_{\mathrm{dimer/step}}-E_{\mathrm{2C/flat}}$&-0.44&-0.33&-0.46&-0.15&-2.54&-2.18\\
\end{tabular}
\end{ruledtabular}
\end{table}

	Now that the nucleation sites on Cu(111) have been identified, we next show that on Ir(111) and Ru(0001) nucleation can be readily facilitated by the step edges. Table~\ref{table1} compares the binding energies of a C dimer at lower step edges with the cases of a dimer on flat surfaces and two separate C adatoms on flat surfaces, showing that dimers at step edges are not only much more stable than on flat surfaces, but also more stable than two separate C adatoms. Therefore, even though on flat Ir and Ru surfaces C dimerization is not preferred, C adatoms attract each other at lower edges of the surface steps. To better illustrate this, we plot the binding energies of two C adatoms on stepped metal surfaces with their separation in Fig.~\ref{figure3}. In all cases, there is a deep potential well upon the formation of a C dimer at lower step edges.

	Summarizing the above results, we have shown that on Ir(111) and Ru(0001), nucleation of C adatoms first occurs at substrate step edges, in agreement with existing experiments \cite{corauxnjp09,loginovanjp08}; whereas on Cu(111), our results predict that C adatoms should nucleate everywhere on the surface. We note that even though there is also a deeper potential well for the C dimer formation at the Cu(111) steps, such steps are not so crucial in the nucleation of C adatoms on Cu, because C adatoms are already strongly attractive to each other on the terraces and readily form dimers before they can reach a step edge. For the same reason we ignored the discussion about Cu(111) steps in the study of single C adatoms above.

	For C adatoms on Ir(111) and Ru(0001), this exceptional tendency towards dimerization at substrate step edges is related to the special local bonding geometry of a C dimer at those sites. In Figs.~\ref{figure3} (a) and (b) the bonding geometries of a C dimer at A-type and B-type step edges are shown, respectively. By comparing those with a C dimer on flat surfaces shown in the inset of Fig.~\ref{figure2}, one can observe that the bonds in the latter case are severely twisted. Since the covalent bonds are highly directional and it is energetically costly to change the relative bond angles, the relaxation of the covalent bonds by the step geometry leads to the extra stability of the C dimers.

	The contrasting behavior of the interacting C adatoms on flat close-packed Ir(111), Ru(0001), and Cu(111) surfaces can be attributed to the competition between the C-C and C-metal interactions. The C-C bond lengths of carbon dimers on flat Ir(111), Ru(0001) and Cu(111) surfaces are 1.397 {\AA}, 1.376 {\AA}, and 1.299 {\AA}, respectively, which are very close to the length of a C-C double bond (1.34 {\AA}). A double bond requires two bonding electrons from each C adatom, but one carbon adatom has only four valence electrons and three nearest metal neighbors on the surface. So intuitively, the formation of a C dimer will weaken the C-metal bonding because of less bonding electrons. Therefore, if the C-metal bonds are very strong, which is the case of Ir and Ru, the dimer formation is not energetically favorable. Conversely, in the case of Cu where C-metal bonding is weak, formation of a dimer is preferred for two C adatoms.

\begin{figure}
 \begin{center}
\includegraphics[width=2.8 in]{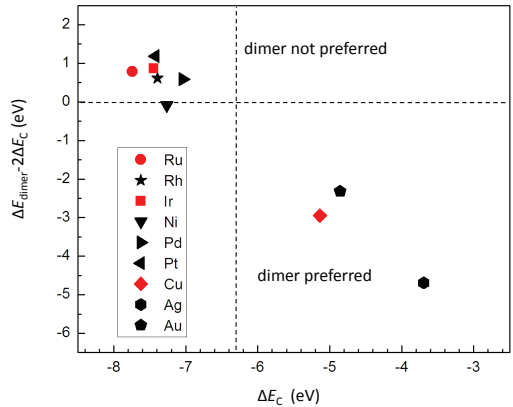}
 \end{center}
 \caption{\textbf{(Color online)}. Binding energy difference between a C dimer ($\triangle E_{\mathrm{dimer}}$) and two separate C adatoms ($2\triangle E_{\mathrm{C}}$) with respect to $\triangle E_{\mathrm{C}}$ on close-packed transition metal surfaces. Vertical dashed line corresponds to the binding energy of a C-C double bond (-6.33 eV) \cite{wade06}.}
 \label{figure4}
\end{figure}

	Next we show that the above picture is not limited to the three representative cases, but can be generalized into a simple guiding principle. To this end, we compare the binding energies of C adatoms and C dimers on the close-packed surfaces of various transition metals, as shown in Fig.~\ref{figure4}. It is apparent that the weaker the C-metal interaction is, the more preferred the C dimers are. In all the cases of noble metals, which have closed \emph{d}-shells and strong free-electron like surface states, C dimerization is preferred. The relative strength of C-metal and C-C interactions largely determines whether the net interaction between C adatoms is attractive or repulsive. The dimer-preferred and dimer-not-preferred systems are essentially separated by the vertical dashed line corresponding to the energy of a C-C double bond (-6.33 eV) \cite{wade06}. The deviation from this trend may be, for example, because of the variation in bonding nature or geometrical effects. Based on the results presented earlier and the prototypical nature of the systems we have studied therein, we can further conclude that for those systems in which C dimers are not preferred on terraces, C nucleation should first occur at substrate step edges. Thus, our study makes it possible to predict where the initial nucleation should happen armed solely with the knowledge of the binding energy of C adatoms to the metal substrate.

This generic principle can lead to many strong predictions. For example, in the strong C-metal binding regime, a flat substrate with scarce steps may not result in growth of quality graphene because of the simultaneous nucleation at multiple sites on the terraces \cite{loginovanjp08}, a somewhat counterintuitive conclusion. In the weak C-metal binding regime, epitaxy on single-crystal flat Cu(111) is more likely to yield graphene with the desired high quality and potential mass production, because C adatoms prefer to nucleate everywhere, and the mismatch of graphene with Cu substrate is very small.

	We finally emphasize on another salient prediction of the present study. The most stable configuration of a C dimer at A- or B-type step edges are shown in Figs.\ref{figure3} (a) and (b), with the dimers bridging the step ridge and the lower terrace perpendicular to the step. Experimentally it is already possible to prepare near-perfect vicinal surfaces containing regular straight step arrays by precise control of the miscut \cite{guoprb06}. Therefore on Ir(111), Ru(0001), and other metal surfaces with strong C-metal bonding, growth of straight graphene nanoribbons with zigzag edges are likely to be obtained from the regularly spaced C dimer arrays formed along the A- or B-type steps in the nucleation stage. This suggestion may offer a potentially more attractive route for mass production of highly ordered graphene nanoribbons with zigzag edges than the known approaches \cite{jiascience09,camposnl09}.

	In summary, we have performed a comparative study of the energetics and kinetics of carbon adatoms on stepped Ir(111), Ru(0001) and Cu(111) surfaces, with intriguing predictions. We have found that the substrate step edges cannot effectively capture single C adatoms, and two carbon adatoms dislike forming dimers on flat Ir(111) and Ru(0001) surfaces either, though they strongly attract each other on Cu(111). However, the lower edges of steps on Ir(111) and Ru(0001) can readily mediate the formation of C dimers. These findings have been rationalized by considering the competition between the carbon-carbon and carbon-metal bonds. We have further generalized this picture to a simple guiding principle for predicting the initial nucleation sites of C adatoms on many other metal substrates, and the predicted behaviors in the early stages of epitaxial graphene growth are expected to be instrumental in achieving mass production of high quality epitaxial graphene.

	The authors thank Brandon Bell for a critical reading of the manuscript. This work was supported by the Division of Materials Science and Engineering, Office of Basic Energy Sciences, Department of Energy, and in part by NSF grant number 0906025. The calculations were performed at NERSC of DOE.

\bibliographystyle{unsrt}

\end{document}